\documentstyle[aps]{revtex}
\newcommand{\beq}{\begin{equation}}
\newcommand{\eeq}{\end{equation}}

\newcommand{\id}
 {i\kern.06em\hbox{\raise.25ex\hbox{$/$}\kern-.60em$\partial$}}

\newcommand{\bs}{/\kern-.52em b}
\newcommand{\qs}{/\kern-.52em s}

\newcommand{\yp}{^{\prime}}

\newcommand{\dd}
{\kern.06em\hbox{\raise.25ex\hbox{$/$}\kern-.60em$\partial$}}

\newcommand{\no}{\nonumber}
\begin{document}
\draft
\tighten
\twocolumn[\hsize\textwidth\columnwidth\hsize\csname@twocolumnfalse\endcsname
\title{New superconducting states in the Hubbard model  }
\author{Sze-Shiang Feng$^{\dag 1,2}$,  Bin Wang$^{\dag 1}$,  Elcio
Abdalla$^{\dag 1}$}
\address{ 1. Instituto De Fisica, Universidade De Sao Paulo, C.P.66.318, CEP
05315-970, \\ Sao Paulo, Brazil \\
2. Department of Astronomy and Applied
  Physics, University of Science and\\ Technology of China, 230026, Hefei,China
   }
\date{\today}
\maketitle
\baselineskip 0.3in
\begin{abstract}
By acting on the excited states at half-filling in the
positive-$U$ Hubbard model with the $\eta$-pairing operators, we
found some new superconducting states in the Hubbard model away
from half-filling. Such eigenstates are different from the
zero-pair and one-pair states introduced by Yang. It is shown
 that the one-pair states are
 not the ground-states and possess ODLRO.
\end{abstract}
]
The explanation of the high temperature superconductivity (HTSC) in doped
cuprates\cite{s1} still remains a challenge for theoretical
physics. Nevertheless, some aspects of the
underlying physics are almost clear: the charge carriers must be
strongly correlated, since at half-filling, the cuprates are
antiferromagnetic insulators, and this can not be explained
definitely by energy -band theories. The system should be
different from the conventional Fermi liquids in the normal
phase. The charge carriers are also paired and the pair is of the
$d$-wave symmetry\cite{s2}. As the simplest model for strongly
correlated electrons, the Hubbard model is believed to
explain the HTSC because it can presumably incorporate the above
key features of the physics\cite{s3}. If this is correct, the
model away from half-filling should display superconductivity at
low temperatures. Especially, the ground-states should be
superconducting. For a negative coupling, Yang first
showed for a modified Hubbard model\cite{s4} that there exist a number of
states (even at
half-filling) which possess off-diagonal long range
order(ODLRO)\cite{s5} and thus are superconducting
 (the $\eta$-pairing states are still superconducting
if the sign of $W$ is changed). The $\eta$-pairing
operators are just the pseudo-spin operators introduced in
reference \cite{s6}. For the generic negative-$U$ Hubbard model,
it was also proved that for sufficiently large $|U|$\cite{s7} or
a special lattice configuration\cite{s8}, and filling, the
ground-state does possess ODLRO. But, experiments indicate that
the cuprates are Mott insulator at half-filling. Therefore, the
Hubbard model coupling should be positive. Hence,
 we should study whether there exist superconducting states for
  the positive-$U$ Hubbard model, especially away from half-filling. 

In this Letter, we first use the pseudo-spin operators to
 generate states away from half-filling from excited states at
 half-filling.
 We will show that these states are different from the
 $\eta$-pairing states introduced by Yang but they also possess ODLRO.
  Then we will prove that the one-pair states also possess ODLRO,
 although  not being the ground-states.
We  first give some preliminaries. The generic Hubbard model is defined
by the Hamiltonian 
\begin{equation}
H=\sum_{<jk>\sigma}t_{jk}c^\dag_{j\sigma}c_{k\sigma}+U\sum_j
n_{j\uparrow}n_{j\downarrow}\quad , \label{hamiltonian}
\end{equation} 
where $c_{j\sigma},c_{j\sigma}^\dag$ are the annihilation and creation
operators for electrons at site $j$ with spin $\sigma$ and $<jk>$
means $j,k$ are nearest neighboring sites (therefore the lattice is
bipartite in the sense of Lieb\cite{s9}). For simplicity, we consider
the 2 dimensional square lattice $\Lambda$ with $N_\Lambda$ lattice
sites.  We denote
lattice vectors ${\bf e}_x,{\bf e}_y$. Then the generalized pseudospin
operators can be defined as (here, the meaning of
  $(-1)^j$ is the same as $e^{i{\bf Q}\cdot{\bf R}_j}$ where ${\bf
  Q}:=(\pi,\pi)$).
\begin{eqnarray}
\eta^\dag_a &=&\sum_j(-1)^jc^\dag_{j\downarrow}c^\dag_{j+a\uparrow}\; ,\quad
\eta_a=\sum_j(-1)^jc_{j+a\uparrow}c_{j\downarrow} \; ,\quad\nonumber\\ 
\eta_z&:=&\frac{1}{2}[\eta_a,
\eta^\dag_a]=\frac{1}{2}(N_\Lambda-\hat{N}_\uparrow-
\hat{N}_\downarrow) 
\end{eqnarray}
where ${\bf a}=m{\bf e}_x+n{\bf e}_y, m$ and $n$ are integers. 
$\hat{N}_\sigma$ is the total number
operator of the electrons with spin $\sigma$. The pseudo-spin
operators $\tilde{\bf S}$ are related to the $\eta$-pairing
operators as \beq \tilde{S}^+:=\eta,\,\,\,\,
\tilde{S}^-:=\eta^\dag,\,\,\,\,\, \tilde{S}_z:=\eta_z \eeq The
total spin operator ${\bf S}$ and the total momentum are defined as 
\begin{eqnarray}
S^\dag&:=&\sum_jc^\dag_{j\uparrow}c_{j\downarrow},
S^-:=\sum_jc^\dag_{j\downarrow}c_{j\uparrow},
S_z:=\frac{1}{2}(\hat{N}_\uparrow-\hat{N}_\downarrow) \nonumber\\
 \hat{\bf P}&=&\sum_{\bf k}{\bf k}(c^\dag_{{\bf
k}\uparrow} c_{{\bf k}\uparrow}+ c^\dag_{{\bf k}\downarrow}
c_{{\bf k}\downarrow})=\hat{\bf P}_\uparrow+\hat{\bf P}_\downarrow\; ,
 \end{eqnarray}
where $c_{i\sigma}=N^{-1/2}_\Lambda\sum_{\bf k}e^{i{\bf
k}\cdot{\bf R}_i}c_{{\bf k}\sigma}$. The momentum  commutes with the spin and
pseudospin. Since the Hamiltonian (\ref{hamiltonian}) commutes with
the set $({\bf P},{\bf S},\tilde{\bf S})$ the generic model therefore enjoys
SU(2)$\otimes$U(1)$\otimes$U(1) symmetry and the
eigenstates of the Hamiltonian can be designated by $|E,{\bf
P};s,s_z;\tilde{s},\tilde{s}_z>$, where $E,{\bf
P};s,s_z;\tilde{s},\tilde{s}_z$ are the cooresponding eigenvalues
of the operators $H, \hat{\bf P}, {\bf S}^2, S_z, \tilde{\bf
S}^2, \tilde{S}_z$. In particular $|E,{\bf P};s,s_z;\tilde{s},0>$
denote the states at half-filling, and the $\tilde{s}$
 must be an integer. According to Lieb\cite{s9}, $s=0$ for the ground-state
and it can also be shown that $\tilde{s}=0$\cite{s8}\cite{s9} is the unique
 ground-state. The excited states can be denoted by $|E,{\bf
   P};s,s_z;\tilde{s},0>$  with $s\not=0, \tilde{s}\not=0$.
The partial particle-hole transformation $\hat{T}$ is defined by 
\beq
\hat{T}c_{i\uparrow}\hat{T}^{-1}=(-1)^ic_{i\uparrow}^\dag,\,\,\,\,
\hat{T}c_{i\downarrow}\hat{T}^{-1}=c_{i\downarrow} 
\eeq
and implements the transformations $\hat{T}{\bf S
  }\hat{T}^{-1}=\tilde{\bf S}$, and
$\hat{T}\hat{\bf P }\hat{T}^{-1}=\hat{\bf P}-N_\uparrow{\bf Q}$.

The Hamiltonian (\ref{hamiltonian}) is thus transformed into 
\beq
\hat{T}H(U)\hat{T}^{-1}=H(-U)-U(S_z+\tilde{S}_z)+\frac{1}{2}UN_\Lambda
\eeq 
The state $|E,{\bf P};s,s_z;\tilde{s},0>$ is transformed into
the state $|E,\hat{\bf P}-N_\uparrow{\bf Q};\tilde{s},0;s,s_z>$. 
The state $|E,\hat{\bf P}-N_\uparrow{\bf Q};\tilde{s},0;s,s_z>$ accommodates
$N_\Lambda-2s_z$ electrons, and the total spin, now equal
to $\tilde{s}$ satisfies $\tilde{s}\le\frac{1}{2}N_\Lambda-s_z$.
Since the $z$-component of the spin satisfies $-\frac{1}{2}N_\Lambda\le
s_z\le \frac{1}{2}N_\Lambda$, we arrive at the following
lemma.

{\it Lemma}: The total pseudo-spin $\tilde{s}$ of the half-filling
states $|E,{\bf P };s,s_z;\tilde{s},0>$ for the model (1) 
satisfies the constraint $\tilde{s}\le N_\Lambda$.
\\
Since $ [\eta^\dag_a, H]=-U\eta^\dag_a$ for $ {\bf a}=0$, while
\begin{eqnarray*}
[\eta^\dag_a, H] =
-U\sum_j(-1)^jc^\dag_{j\downarrow}c^\dag_{j+a\uparrow}
(n_{j\uparrow}+n_{j+a\downarrow}) ,
\end{eqnarray*}
for ${\bf a}\not=0 $,  according to Yang\cite{s4}, the  states  $|Y_{n,a}>
:=(\eta^\dag)^{n-1}\eta^\dag_a|0>$ and $|Y_n>:=(\eta^\dag)^n|0>$
are $2n$-electron eigenstates for all ${\bf a}$. Our
first result can thus be stated as

{\it Theorem 1}: Suppose $|E,{\bf P};s,s_z;\tilde{s},0>$ is an
excited state at half-filling, then the states $
|\Phi^-_n>:=\eta^n |E,{\bf P};s,s_z;\tilde{s},0> $ and $
|\Phi^+_n>:=(\eta^{+})^n|E,{\bf P};s,s_z;\tilde{s},0> $,\,\,\,\,\,
 for $ n<{\rm min}(\tilde{s},\frac{N_\Lambda}{2}) $, possess ODLRO
and are different from $|Y_{\frac{N_\Lambda}{2}-n,a}> $ and
$|Y_{\frac{N_\Lambda}{2}-n}>$ for $\tilde{s}\not=\frac{N_\Lambda}{2}$.

{\it Proof}: First let us consider the state $ |\Phi^-_n>$. It
contains $N_\Lambda-2n$ electrons and we must show that they do not
vanish. Using the angular-momentum commutation relations
we have
\beq \tilde{\bf
S}^2-\tilde{S}_z^2-\tilde{S}_z=\tilde{S}^-\tilde{S}^+ ,\eeq
Therefore, \beq <\Phi^-_1|\Phi^-_1>=<\Phi_0|\tilde{\bf
S}^2-\tilde{S}_z^2-\tilde{S}_z|\Phi_0>
 =\tilde{s}(\tilde{s}+1)
\eeq
Similarly, we have
\begin{eqnarray}
<\Phi^-_n|\Phi^-_n>&=&<\Phi^-_{n-1}|\tilde{\bf S}^2-\tilde{S}_z^2-
\tilde{S}_z|\Phi^-_{n-1}>\no\\
&=&[\tilde{s}(\tilde{s}+1)
-n(n-1)]<\Phi^-_{n-1}|\Phi^-_{n-1}> ,
\end{eqnarray}
Therefore, if $|\Phi_0>$ is normalized, we have (similarly for $|\Phi^+_n>$
\beq
<\Phi^\mp_n|\Phi^\mp_n>=\prod_{l=1}^n[\tilde{s}(\tilde{s}+1)-l(l-1)]
\eeq 
Therefore, $|\Phi^\pm_n>\not=0$. Since \beq
H|\Phi^\pm_n>=(E\pm nU)|\Phi_n> \eeq $|\Phi^\pm_n>$ are also
eigenstates. Now let us investigate the ODLRO in $|\Phi^\pm_n>$,
i.e.
$\lim_{|i-j|\rightarrow\infty}<\Phi^\pm_n|\eta^\dag_i\eta_j|\Phi^\pm_n>$.
For this purpose, we investigate the quantity
$<\Phi^\pm_n|\tilde{\bf S}^2- \tilde{S}_z^2|\Phi^\pm_n>$,
 \begin{eqnarray}
<\tilde{\bf S}^2-\tilde{S}_z^2>&=&\frac{<\Phi^\pm_n|\tilde{\bf S}^2-
\tilde{S}_z^2|\Phi^\pm_n>}{<\Phi^\pm_n|\Phi^\pm_n>}=\tilde{s}(\tilde{s}+1)-n^2
\no\\&=&
\beta^2(\frac{1}{\alpha^2}-1)N_\Lambda^2+\frac{\beta}{\alpha}N_\Lambda
\end{eqnarray} 
where we used that $n=\alpha\tilde{s}=\beta N_\Lambda$.
According to a theorem by Yang and Zhang\cite{s10}, there
exists ODLRO in $|\Phi_{\beta N_\Lambda}>$ so long as $\alpha<1$.
Obviously, $|\Phi_0>=|E; s,s_z; \frac{\beta}{\alpha}
N_\Lambda,0>$ also possess ODLRO. 

Next we show that $|\Phi^\pm_n>$ are different from
$|Y_{\frac{N_\Lambda}{2}-n}>$ and $|Y_{\frac{N_\Lambda}{2}-n,a}>$
for $\tilde{s}\not=\frac{N_\Lambda}{2}$. Since 
the eigenvalue of ${\bf S}^2 $ corresponding to $|\Phi^\pm_n>$ is 
$\tilde{s}(\tilde{s}+1) $ while the one corresponding to $|Y_{\frac{N_
\Lambda}{2}-n}> $ is $\frac{N_\Lambda}{2} (\frac{N_\Lambda}{2}+1)$,
the eigenvectors must be mutually orthogonal.
Using 
\begin{eqnarray*}
[\eta,
\eta^\dag_a]=\sum_j[c_{j\uparrow}c_{j+a\uparrow}^\dag-(-1)^a
 c_{j\downarrow}c_{j+a\downarrow}^\dag] ,\;
[\eta^z,\eta^\dag_a]=-\eta^\dag_a ,
\end{eqnarray*}
we have
\begin{eqnarray}
\tilde{\bf S}^2|Y_{\frac{N_\Lambda}{2}-n,a}>
&&=(\eta^\dag)^{\frac{N_\Lambda}{2}-n-1}([\tilde{\bf
S}^2,\eta^\dag_a]+\eta^\dag_a\tilde{\bf S}^2)|0>\no\\
=&&(\eta^\dag)^{\frac{N_\Lambda}{2}-n-1} ([
\eta^\dag\eta+\eta_z^2+\eta_z,\eta^\dag_a]
+\eta^\dag_a\tilde{\bf S}^2)|0>\no\\
=&&\frac{N_\Lambda}{2}(\frac{N_\Lambda}{2}-1)|Y_{\frac{N_\Lambda}{2}-n,a}>
\end{eqnarray}
Hence $|\Phi^\pm_n>$ is different from
$|Y_{\frac{N_\Lambda}{2}-n,a}>$. This completes our proof of {\it
theorem 1}.

We know that for positive $U$, the energy of
$|Y_{\frac{N_\Lambda}{2}-n,a}>$
  is lower than that of
$|Y_{\frac{N_\Lambda}{2}-n}>$ by $U$: $E_{\frac{N_\Lambda}{2}-n}-
E_{\frac{N_\Lambda}{2}-n,a}=U$ ( but in the thermodynamic limit,
the energy-densities of these two states are equivalent).  In
\cite{s1}, Yang proved that $|Y_{\frac{N_\Lambda}{2}-n}>$ is not
the ground-state by showing that the energy of
$|Y_{\frac{N_\Lambda}{2}-n}>$ is equal to the energy expectation
of the (non-eigen)state $|\phi_1>:
=(\eta_C^\dag)^{\frac{N_\Lambda}{2}-n}|0>$ where $\eta_C$ is the
Cooper pair, $\eta_C:=\sum_jc_{j\uparrow}c_{j\downarrow}$. Our
second result concerns $|Y_{\frac{N_\Lambda}{2}-n,a}>$ and can be
stated as

{\it Theorem 2}: The state $|Y_{\frac{N_\Lambda}{2}-n,a}>$ is not a
ground state for $N_\Lambda-2n$ electrons and possesses
ODLRO.

{\it Proof}: We first show that it is not a ground-state. For this
purpose, we consider the state 
\beq
|\phi_2>:=(\eta^\dag_C)^{\frac{N_\Lambda}{2}-n-1}\eta^\dag_a|0> 
\label{statenot}
\eeq
and the energy expectation value in it. From \break
$ [U\sum_j n_{j\uparrow}n_{j\downarrow}, \eta^\dag_C]=U\eta^\dag_C$
we have $[U\sum_j n_{j\uparrow}n_{j\downarrow},
(\eta^\dag_C)^m]=mU(\eta^\dag_C)^m$, therefore 
\beq
\frac{<\phi_2|U\sum_j
n_{j\uparrow}n_{j\downarrow}|\phi_2>}{<\phi_2|\phi_2>}
=U(\frac{N_\Lambda}{2}-n-1)\quad . 
\eeq 
Now consider the kinetic
energy:  $T=\sum_{jk,\sigma}t_{jk}c^\dag_{j\sigma}c_{k\sigma}$.
In ${\bf k}$-space, 
($m=\frac{N_\Lambda}{2}-n-1$)
\begin{eqnarray*}
|\phi_2>=\!\sum_{{\bf p},{\bf k}_1...{\bf k}_m }
\! e^{i{\bf p}\cdot{\bf a}}
c^\dag_{-{\bf k}_1\downarrow}
c^\dag_{{\bf k}_1\uparrow} ... c^\dag_{-{\bf k}_m\downarrow}c^\dag_{{\bf
    k}_m\uparrow} c^\dag_{{\bf Q}- {\bf p}\downarrow}c^\dag_{{\bf
    p}\uparrow}|0> 
\end{eqnarray*}
Obviously, the ${\bf k}$'s in the sum $\sum_{{\bf k}_1,{\bf k}_2,...
{\bf k}_m}$ differ from each other. Now consider the ${\bf k}_1$-pair.
For a given ${\bf p}\in BZ$, the ${\bf k}_1$-pair contributes
when ${\bf k}_1\not={\bf p}, {\bf k}_1\not={\bf p}-{\bf Q}$.
For each such ${\bf k}_1$ corresponds ${\bf k}\yp_1:={\bf Q}-{\bf  k}_1$; 
the ${\bf k}\yp_1$-pair also makes contributions
  when ${\bf k}\yp_1\not={\bf p}, {\bf k}\yp_1\not={\bf p}-{\bf Q}$,
  thus, both ${\bf k}_1$- and the ${\bf k}\yp_1$-pair contribute
simultaneously when
\beq 
{\bf k}_1\not={\bf p}, \pm({\bf p}-{\bf Q}), 2{\bf Q}-{\bf
p}\simeq -{\bf p} \quad .
\eeq 
The energy of the ${\bf k}_1$-pair and
the corresponding ${\bf k}\yp_1$-pair cancel each other because of $t({\bf
k})=-t({\bf Q}-{\bf k})$. Thus only when \beq {\bf k}_1=-{\bf p},
{\rm or} \,\,\, {\bf k}_1={\bf Q}-{\bf p} \eeq the ${\bf
k}_1$-pair will contribute to the kinetic energy. Since the other
${\bf k}$'s are different from ${\bf k}_1$, the corresponding
pairs contribute
 nothing to the kinetic energy, and it is the same as
\begin{eqnarray*}
\frac{<\phi_2|T|\phi_2>}{<\phi_2|\phi_2>}=m\sum_{\bf p}
 (\frac{<\varphi_1|T|\varphi_1>}{<\varphi_1|\varphi_1>}
 +\frac{<\varphi_2|T|\varphi_2>}{<\varphi_2|\varphi_2>})
\end{eqnarray*}
where $|\varphi_1>=|c^\dag_{-({\bf Q}-{\bf p})\downarrow}c^\dag_{{\bf Q}-{\bf
p}\uparrow} c^\dag_{{\bf Q}-{\bf p}\downarrow}c^\dag_{{\bf p}\uparrow}>$
and
$ |\varphi_2>=|c^\dag_{{\bf p}\downarrow}c^\dag_{-{\bf p}\uparrow}
c^\dag_{{\bf Q}-{\bf p}\downarrow}c^\dag_{{\bf p}\uparrow}>$.
\\
Therefore
\begin{eqnarray}
\frac{<\phi_2|T|\phi_2>}{<\phi_2|\phi_2>} &&=
m\sum_{\bf p}[t({\bf p}-{\bf Q})+t({\bf Q}-{\bf p})+ t({\bf Q}-{\bf
p})\no\\
+&&t({\bf p})+t({\bf p})+t(-{\bf p})+t({\bf Q}-{\bf p})+t({\bf
p})] =0
\end{eqnarray}
where we have used that $t({\bf k})=t(-{\bf k}), t({\bf
k})=-t({\bf Q}-{\bf k})$. Finally, 
\beq
\frac{<\phi_2|H|\phi_2>}{<\phi_2|\phi_2>}=U(\frac{N_\Lambda}{2}-n-1)
\eeq 
i.e., $|\phi_2>$ has the same energy as that of
$|Y_{\frac{N_\Lambda}{2}-n,a}>$, but $|\phi_2>$ is not an
eigenstate, hence $|Y_{\frac{N_\Lambda}{2}-n,a}>$ is not
the ground-state.

The next is to show that it possesses ODLRO.  For this purpose, we
have to know the norm of $|Y_{\frac{N_\Lambda}{2}-n,a}>$. Define
\beq \beta_n:=<0|\eta_a\eta^n(\eta^\dag)^n\eta_a^\dag|0> \eeq For
$n=1$, direct calculation gives $\beta_1=N_\Lambda(N_\Lambda-2)$.
For $\beta_2$, we can use the relation $[\eta,\eta^\dag]=2\eta_z$
to obtain 
\beq
\beta_2=2(N_\Lambda-3)\beta_1+<0|\eta_a\eta\eta^\dag\eta^\dag\eta\eta^\dag_a|0>
\eeq 
Using eq. (\ref{statenot}), the second term vanishes. Therefore
$\beta_2=2N_\Lambda(N_\Lambda-2)(N_\Lambda-3)$. In general, one
can obtain the relation \beq \beta_n=n(N_\Lambda-n-1)\beta_{n-1}
\eeq 
Thus we have 
\beq
\beta_n=\frac{n!}{N_\Lambda-1}\frac{N_\Lambda!}{(N_\Lambda-n-2)!}
\eeq
Therefore, we should have $n\le N_\Lambda-2$ for
$|Y_{n+1,a}>$.
Now let us check the behaviour of
$C_n:=<Y_{n,a}|C_{r\uparrow}C_{r\downarrow}C^\dag_{s\downarrow}
C^\dag_{s\uparrow}|Y_{n,a}>$
as $|r-s|\rightarrow\infty$. We define 
\begin{eqnarray*}
D_n&:=&<0|\eta_a
C_{r\uparrow}C_{r\downarrow}\eta^{n-1}(\eta^\dag)^n\eta
C^\dag_{s\downarrow}C^\dag_{s\uparrow}\eta^\dag_a|0> \\
E_n&:=&<0|\eta_a
C_{r\uparrow}C_{r\downarrow}\eta^{n-3}\eta\eta^\dag(\eta^\dag)^{n-1}(\eta^2
C^\dag_{s\downarrow}C^\dag_{s\uparrow}\eta^\dag_a)|0> 
\end{eqnarray*}
We can obtain the recurrence relations 
\begin{eqnarray*}
E_n&=&n(N_\Lambda+1-n)E_{n-1}\\
D_n&=&n(N_\Lambda-n-1)D_{n-1}+E_n \\
C_n&=&n(N_\Lambda-n-3)C_{n-1}+D_n
\end{eqnarray*}
Since $ \eta^2C^\dag_{s\downarrow}C^\dag_{s\uparrow}\eta^\dag_a|0>=0 $
we have $E_n=0$, hence 
\beq 
D_n=n(N_\Lambda-n-1)D_{n-1}
\eeq 
By direct calculation, we know $D_1=(N_\Lambda-4)e^{i{\bf
Q}\cdot({\bf R_r}-{\bf R_s})}$, therefore 
\beq
D_n=n!(N_\Lambda-4)\frac{(N_\Lambda-3)!}{(N_\Lambda-n-2)!}e^{i{\bf
Q}\cdot({\bf R_r}-{\bf R_s})} \eeq Accordingly, we can obtain
 \begin{eqnarray}
C_n&=&n!\frac{(N_\Lambda-5)!}{(N_\Lambda-4-n)!}C_1\no\\&&+D_n[1
+(\frac{N_\Lambda-3-n}{N_\Lambda-1-n})\no\\&&
+(\frac{N_\Lambda-3-n}{N_\Lambda-1-n}\cdot
\frac{N_\Lambda-2-n}{N_\Lambda-n})+...+\no\\&&
(\frac{N_\Lambda-3-n}{N_\Lambda-1-n}\cdot\frac{N_\Lambda-2-n}{N_\Lambda-n}...
\frac{N_\Lambda-4}{N_\Lambda-2})]
\end{eqnarray}
Therefore \beq
\lim_{|r-s|\rightarrow\infty}C_n=n!\frac{(N_\Lambda-5)!}{(N_\Lambda-4-n)!}C_1+(n-1)D_n
\eeq Since $C_1=(N_\Lambda-4)e^{i{\bf Q}\cdot({\bf R_r}-{\bf
R_s})}$, we have \begin{eqnarray}
\lim_{|r-s|\rightarrow\infty}C_n&=&e^{i{\bf Q}\cdot({\bf
R_r}-{\bf R_s})}[n!\frac{(N_\Lambda-4)!}{(N_\Lambda-4-n)!}
\no\\&&+n!(n-1)\frac{(N_\Lambda-4)(N_\Lambda-3)!}{(N_\Lambda-n-2)!}]
 \end{eqnarray}
 Finally, we arrive at
 \begin{eqnarray}
\lim_{|r-s|\rightarrow\infty}\frac{<Y_{n,a}|C_{r\uparrow}
C_{r\downarrow}C^\dag_{s\downarrow}C^\dag_{s\uparrow}|Y_{n,a}>}
{<Y_{n,a}|Y_{n,a}>}
=e^{i{\bf Q}\cdot({\bf R_r}-{\bf
R_s})}\frac{(n-1)(N_\Lambda-4)}{N_\Lambda(N_\Lambda-2)}
\end{eqnarray}
This completes our proof of {\it Theorem 2}.

We now pass to the conclusions. First, when $\tilde{s}\not=0$, at
least one site is double-occupied in the half-filling state $|E,{\bf P
};s,s_z;\tilde{s},0>$ because $\eta|E,{\bf P
};s,s_z;\tilde{s},0>\not=0$. Therefore, all excited states at
half-filling have at least one site double-occupied. Second, we
have obtained the explicit expression of $
\lim_{|r-s|\rightarrow\infty} <\eta_r\eta^\dag_s>$. In fact, to
show the existence of ODLRO, we can also use a theorem from
\cite{s10}. Indeed, since
 \beq
\frac{<Y_n|\tilde{\bf
S}^2-\tilde{S}_z^2|Y_n>}{<Y_n|Y_n>}=\frac{N_\Lambda}{2}+nN_\Lambda-n^2
\eeq
 we also have
\beq 
\frac{<Y_{n,a}|\tilde{\bf
S}^2-\tilde{S}_z^2|Y_{n,a}>}{<Y_{n,a}|Y_{n,a}>}=-
\frac{N_\Lambda}{2}+nN_\Lambda-n^2
\eeq 
which is of order $O(N_\Lambda^2)$ when $n$ is of oder
$N_\Lambda$. Finally, we do not know whether the states
$|\Phi_n>$ are ground-states or not. We do not know either
which one of the states $|\Phi_n>$ and
$|Y_{\frac{N_\Lambda}{2}-n,a}>$ has a lower energy. Though we do
not know whether the ground-states away from half-filling are
superconducting or not, we can make sure that there is a number
of states which are superconducting away from half-filling in the
positive-$U$ Hubbard model. Of course, whether these superconducting
states are responsible for the HTSC remains unclear.

\underline{Acknowledgements:} This work was supported in part by
Funda\c c\~ao de Amparo \`a Pesquisa do Estado de
S\~ao Paulo (FAPESP) and Conselho Nacional de Desenvolvimento
Cient\'{\i}fico e Tecnol\`ogico (CNPQ). S.F. and B.W. 
would also like to acknowledge the support given by 
NNSF, China.

$\dag$e-mails: sshfeng, binwang, eabdalla @fma.if.usp.br


\vspace*{-0.5cm}
\begin{thebibliography}{s40} 
\vspace*{-1.5cm} 
\bibitem{s1} J.G. Bednorz \& K.A. M\"uller {\it Z. Phys.}{\bf B64}
             (1986):189.
\bibitem{s2} C.C. Tsuei \& J.R. Kirtley {\it Rev. Mod. Phys.}{\bf
    72}(2000):969. 
\bibitem{s3} P.W. Anderson {\it The Theory of Superconductivity in the
             High-$T_c$ Cuprates} (Princeton University, Princeton,
           New Jersey,1997) 
\bibitem{s4} C.N. Yang {\it Phys. Rev. Lett.} {\bf 63}(1989):2144.
\bibitem{s5} C.N. Yang {\it Rev. Mod. Phys.} {\bf 34}(1962):694.
\bibitem{s6} C. Castellani,C.D. Castro,D. Feinberg,  \&  J. Ranninger
             {\it Phys. Rev. Lett.} {\bf 43}(1979):1957.
\bibitem{s7} R.R.P. Singh \& R.T Scaletta {\it Phys. Rev. Lett.}
             {\bf 66}(1991):3203.
\bibitem{s8} S.Q. Shen \& Z.M. Qiu {\it Phys. Rev. Lett.} {\bf 71}(1993):4238.
\bibitem{s9} E.H. Lieb {\it Phys. Rev. Lett.} {\bf 62}(1989):1201.
\bibitem{s10} C.N. Yang \& S.C. Zhang {\it Mod.Phys.Lett.}
              {\bf B4}(1990):759.
\end{thebibliography}
\end{document}